\documentclass[twocolumn,showpacs,preprintnumbers,amsmath,amssymb,superscriptaddress,pra,longbibliography]{revtex4-1}

\usepackage{graphicx}
\usepackage{dcolumn}
\usepackage{bm}
\usepackage{graphics}
\usepackage{subfigure}
\usepackage{graphicx}
\usepackage{epsfig}
\usepackage{epstopdf}
\usepackage{xcolor}
\usepackage{mathtools}

\allowdisplaybreaks

\begin{document}

\title{Collective modes of an imbalanced unitary Fermi gas}

\author{Johannes Hofmann}
\email{jbh38@cam.ac.uk}
\affiliation{TCM Group, Cavendish Laboratory, University of Cambridge, Cambridge CB3 0HE, United Kingdom}

\author{Fr\'ed\'eric Chevy}
\affiliation{Laboratoire Kastler Brossel, ENS-PSL Research University, CNRS, UPMC, Coll\`ege de France, 24, rue Lhomond, 75005 Paris}

\author{Olga Goulko}
\affiliation{Department of Physics, University of Massachusetts, Amherst, MA 01003, USA }
\affiliation{Raymond and Beverly Sackler School of Chemistry and School Physics and Astronomy, Tel Aviv University, Tel Aviv 6997801, Israel}

\author{Carlos Lobo}
\affiliation{Mathematical Sciences, University of Southampton, Highfield, Southampton, SO17 1BJ, United Kingdom}

\date{\today}

\begin{abstract}
We study theoretically the collective mode spectrum of a strongly imbalanced two-component unitary Fermi gas in a cigar-shaped trap, where the minority species forms a gas of polarons. We describe the collective breathing mode of the gas in terms of the Fermi liquid kinetic equation taking collisions into account using the method of moments. Our results for the frequency and damping of the longitudinal in-phase breathing mode are in good quantitative agreement with an experiment by Nascimb\`ene {\it et al.} [Phys. Rev. Lett. {\bf 103}, 170402 (2009)] and interpolate between a hydrodynamic and a collisionless regime as the polarization is increased. A separate out-of phase breathing mode, which for a collisionless gas is sensitive to the effective mass of the polaron, however, is strongly damped at finite temperature, whereas the experiment observes a well-defined oscillation.
\end{abstract}

\pacs{}
\maketitle

\section{Introduction}

Landau's Fermi liquid theory accounts for the fact that many normal state Fermi systems behave in a qualitatively similar way to a noninteracting Fermi gas~\cite{landau57a,landau57b,landau59}. The central assumption of the theory is the adiabatic continuity of excitations, meaning that excitations of the interacting system are characterized by the same quantum numbers of spin $\sigma=\uparrow,\downarrow$ and momentum ${\bf p}$ as the noninteracting system~\cite{pines94,landau06,giuliani05}. The robustness of this picture relies on phase space arguments and does not depend on the strength of the interparticle interaction.

Over the past ten years, the two-spin-component Fermi quantum gas has emerged as a new Fermi liquid~\cite{lobo06}. For small polarization $P=(N_\uparrow-N_\downarrow)/(N_\uparrow + N_\downarrow)$ ($N_{\uparrow,\downarrow}$ being the total number of atoms of each species), the ground state is a superfluid~\cite{zwerger16}. As the polarization is increased beyond the Clogston-Chandrasekar limit, there is a first order phase transition to a Fermi liquid where both species coexist~\cite{chevy12,recati12}. In particular, the extreme limit $P\to 1$ describes a single spin-$\downarrow$ quasiparticle  interacting with a majority spin-$\uparrow$ Fermi sea (a ``polaron") characterized by an effective mass $m^*$, energy $E_p = - \alpha E_F$ (where $E_F$ is the Fermi energy of the majority species),  and quasiparticle residue. These parameters have been studied extensively at zero temperature~\cite{chevy06,combescot07,veillette08,combescot08,prokofev08a,prokofev08,punk09,vlietinck13,goulko16}.

There are three ways to measure the polaron parameters. First, through the equation of state~\cite{navon10}. Second, by measuring the radiofrequency spectrum, which has a pronounced quasiparticle peak at the polaron energy with a weight proportional to the quasiparticle residue~\cite{schirotzek09,kohstall12,scazza17}. The third method -- which we are interested in here -- measures the effective mass {\em dynamically} by exciting collective mode oscillations~\cite{nascimbene09}.

The experiment~\cite{nascimbene09} by Nascimb\`ene {\it et al.} studied the collective breathing modes in the longitudinal direction of an elongated harmonic trap as a function of polarization. At low polarization, both spin components oscillate in phase due to the strong coupling between them. At larger polarization, an additional out-of-phase mode was observed. In the $P\to1$ limit its frequency was identified with the collisionless value $2\omega^*_z$, where $\omega^*_z$ is the axial trap frequency renormalized by the interaction of the minority atoms with the majority background~\cite{lobo06}:
\begin{align}
\omega^*_z &= \omega_z \sqrt{\frac{m}{m^*} (1+\alpha)} . \label{eq:collfreq}
\end{align}
Reference~\cite{nascimbene09} obtained the polaron effective mass from Eq.~\eqref{eq:collfreq} after linearly extrapolating the experimental out-of-phase breathing mode frequency to $P=1$. This has resulted in a value at unitarity of $m^*/m=1.17(10)$, in close agreement with theoretical results \cite{combescot08,combescot09,pilati08,prokofev08a,prokofev08,vlietinck13}.

In a subsequent theory paper~\cite{recati10}, Recati and Stringari analyzed the out-of-phase collective mode using a scaling ansatz with mean-field interactions but without collisions, and obtained a frequency behaviour that disagreed with the experiment~\cite{nascimbene09} at lower values of polarization. However, at these polarizations collisions can become important so that a full theoretical description of the experiment is still lacking.

In this paper, we analyze the collective breathing mode spectrum of a Fermi liquid taking into account finite-temperature effects, mean-field interactions and also quasiparticle collisions. The theoretical framework that allows us to do this is the Landau-Boltzmann equation, which we solve using the method of moments. This method has already been successfully applied to study the collective modes of balanced Fermi gases~\cite{riedl08,chiacchiera09,lepers10,chiacchiera11,pantel12,chiacchiera13}. The paper is structured as follows: in Sec.~\ref{sec:coll}, we solve the quasiparticle kinetic equation for a trapped and strongly imbalanced Fermi gas using the single-polaron parameters obtained in~\cite{combescot08}. We obtain the eigenmodes in a trap by expanding  the distribution function in small deviations from equilibrium in a finite-dimensional basis of trial functions. In this way, both the single-particle contribution to the kinetic equation as well as the collision integral can be reduced to a set of linear equations whose eigenvalues determine the collective mode frequencies. We present results for collective modes for the experimental setup of Ref.~\cite{nascimbene09}, and compare with the experimental results.

\section{Collective modes}\label{sec:coll}

In the high-polarization limit of the imbalanced Fermi gas, the minority atoms \mbox{(spin-$\downarrow$)} form a dilute gas of polarons that interact with the majority species \mbox{(spin-$\uparrow$)}. Within Fermi liquid theory, the quasi-classical evolution of the minority and majority distribution function $n_\sigma({\bf r},{\bf p},t)$ is described by the coupled Landau-Boltzmann kinetic equation [setting $\hbar=1$],
\begin{align}
&\biggl[\partial_t + \frac{\partial \varepsilon_\sigma({\bf r}, {\bf p})}{\partial {\bf p}}  \cdot \frac{\partial}{\partial {\bf r}} - \frac{\partial \varepsilon_\sigma({\bf r}, {\bf p})}{\partial {\bf r}} \cdot \frac{\partial}{\partial {\bf p}}\biggr] n_\sigma({\bf r},{\bf p}) \nonumber \\
&\quad= - I_\sigma[n_\uparrow, n_\downarrow], \label{eq:kinetic}
\end{align}
where the distribution functions of each spin state are normalized as
\begin{align}
N_\sigma &= \int d^3r \, n_\sigma({\bf r}) \quad {\rm with} \quad n_\sigma({\bf r}) = \int \frac{d^3p}{(2\pi)^3} n_\sigma({\bf r}, {\bf p}). \label{eq:density}
\end{align}
$\varepsilon_\sigma({\bf r}, {\bf p})$ is the energy of a quasiparticle with spin $\sigma$ and momentum ${\bf p}$ at position ${\bf r}$:
\begin{align}
\varepsilon_\sigma({\bf r}, {\bf p}) &= \frac{p^2}{2m_\sigma} + U_\sigma({\bf r}) + V({\bf r}). \label{eq:excitationenergy}
\end{align}
We take $m_\downarrow=m^*$, the effective polaron mass, and $m_\uparrow=m$, the bare atom mass. The spin-independent harmonic trapping potential with trapping frequencies $\omega_i$ ($i=x,y,z$) is given by
\begin{align}
V({\bf r}) = \sum_{i=x,y,z} \frac{m\omega_i^2}{2} r_i^2 .
\end{align}
$U_\sigma=U_\sigma[n_\uparrow(\mathbf{r}),n_\downarrow(\mathbf{r})]$ are the mean-field interactions experienced by each spin component, which are deduced from the single-polaron parameters at zero temperature. For the minority species, the mean field potential is given by the single-polaron energy $U_\downarrow=-\alpha E_F(\mathbf{r})$, while the majority mean field $U_\uparrow$ is chosen such that the total force acting on the system vanishes,
\begin{align}
U_\downarrow[n_\uparrow,n_\downarrow] &= - \alpha \frac{(6\pi^2)^{2/3}}{2m} n_\uparrow^{2/3}({\bf r}) \label{eq:mfminority} 
\\ \nonumber\\[-2ex]
U_\uparrow[n_\uparrow,n_\downarrow] &= - \frac{2}{3} \alpha \frac{(6\pi^2)^{2/3}}{2m} \frac{n_\downarrow({\bf r})}{n_\uparrow^{1/3}({\bf r})} . \label{eq:mfmajortiy}
\end{align}
The distribution in thermal equilibrium is the Fermi-Dirac distribution with chemical potential $\mu_\sigma$~\cite{landau06,pines94}
\begin{align}
n_\sigma^{\rm eq}({\bf r}, {\bf p}) &= \frac{1}{e^{\beta (\varepsilon_\sigma({\bf r}, {\bf p}) - \mu_\sigma)} + 1} . \label{eq:occ}
\end{align}
Unlike for a noninteracting gas, Eq.~\eqref{eq:occ} is a complicated self-consistent expression since $n_\sigma^{\rm eq}({\bf r}, {\bf p})$ enters $\varepsilon_\sigma({\bf r}, {\bf p})$ through the mean-field potential. The attractive mean-field potential increases the particle density in the trap center.
\begin{figure*}[t!]
\subfigure[]{
\scalebox{0.68}{\includegraphics{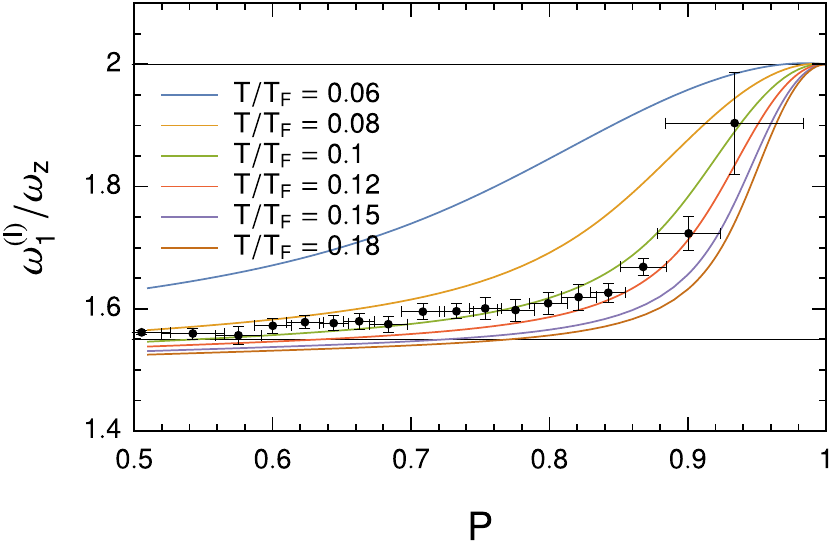}}}
\subfigure[]{
\scalebox{0.69}{\includegraphics{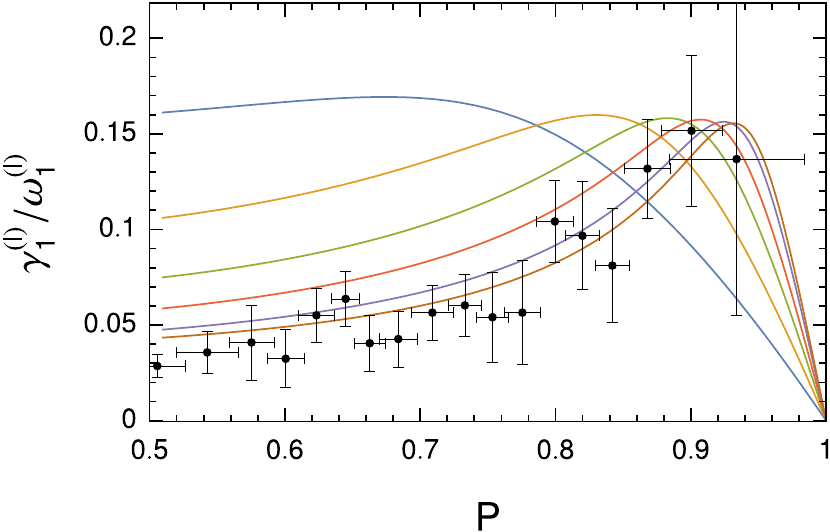}}}
\subfigure[]{
\scalebox{0.67}{\includegraphics{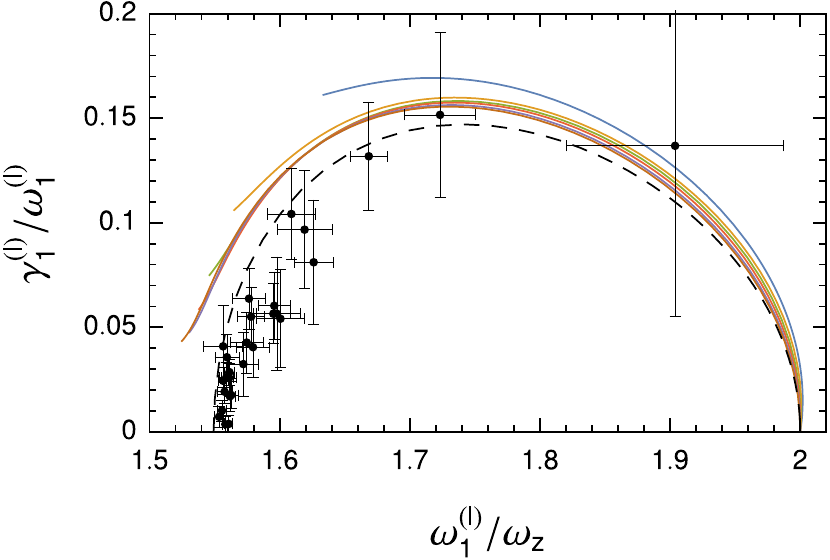}}}
\caption{(a) Frequency and (b) damping of the longitudinal breathing mode of an imbalanced Fermi gas in an anisotropic trap with aspect ratio $\lambda=0.075$ as a function of polarization for different temperatures $T/T_F = 0.06,0.08,0.1,0.12,0.15$ and $0.18$, respectively. The thin black lines in the first figure denote the collisionless limit $\omega_{\rm cl} = 2 \omega_z$ and the hydrodynamic limit $\omega_{\rm hd} = \sqrt{12/5} \omega_z$. (c) Scaling plot of frequency versus damping, which indicates the presence of a single dominant scattering lifetime. The black dashed line is the parametric estimate of Eq.~\eqref{eq:intform}. The experimental data points are taken from Ref.~\cite{nascimbene09}.}
\label{fig:1}
\end{figure*}

Interactions change the distribution function through the collision integral
\begin{widetext}
\begin{align}
I_\sigma[n_\uparrow, n_\downarrow] = \int \frac{d^3p_1}{(2 \pi)^3} \, \int d\Omega \, \frac{d\sigma}{d\Omega} \, |{\bf v}_{\rm rel}| \, &\biggl[ n_\sigma({\bf r},{\bf p}, t) \, n_{\bar{\sigma}}({\bf r},{\bf p}_1, t) \, (1 - n_\sigma({\bf r},{\bf p}', t)) \, (1 - n_{\bar{\sigma}}({\bf r},{\bf p}_1', t)) \nonumber \\
&\qquad- n_\sigma({\bf r},{\bf p}', t) \, n_{\bar{\sigma}}({\bf r},{\bf p}_1', t) \, (1 - n_\sigma({\bf r},{\bf p}, t)) \, (1 - n_{\bar{\sigma}}({\bf r},{\bf p}_1, t)) \biggr] , \label{eq:collision}
\end{align}
\end{widetext}
where $\bar{\sigma}$ denotes the opposite spin species of $\sigma$, ${\bf v}_{\rm rel}$ the relative velocity of colliding particles, and $\frac{d\sigma}{d\Omega} = \bigl(\frac{m_{\rm red}}{2\pi} f_{\uparrow\downarrow}\bigr)^2$ the differential scattering cross section, where the scattering amplitude $f_{\uparrow\downarrow}$ is linked to the single-polaron energy by $f_{\uparrow\downarrow} = \frac{\partial E_p}{\partial n_\uparrow}$. The first line of the collision integral describes the depopulation of the state $({\bf p}, \sigma)$ by collisions with a quasiparticle $({\bf p}_1, \bar{\sigma})$ to a final state $({\bf p}', \sigma)$ and $({\bf p}_1', \bar{\sigma})$. The second line describes the reverse process $({\bf p}', \sigma) + ({\bf p}_1', \bar{\sigma}) \to ({\bf p}, \sigma) + ({\bf p}_1, \bar{\sigma})$. The collisions are constrained by energy and momentum conservation. Writing in the center-of-mass frame ${\bf p} = \frac{m_\sigma}{M} {\bf P} + {\bf q}$ and ${\bf p}_1 = \frac{m_{\bar{\sigma}}}{M} {\bf P} - {\bf q}$, where $M = m + m^*$ is the total mass and ${\bf P} = {\bf p} + {\bf p}_1$ the total momentum, we have ${\bf p}' = \frac{m_\sigma}{M} {\bf P} + {\bf q}'$ and ${\bf p}_1' = \frac{m_{\bar{\sigma}}}{M} {\bf P} - {\bf q}'$, where $|{\bf q}| = |{\bf q}'|$, as well as ${\bf v}_{\rm rel} = {\bf q}/m_{\rm red}$ with $m_{\rm red} = \frac{mm^*}{m+m^*}$ the reduced mass. The integration over the angle element $d\Omega$ in Eq.~\eqref{eq:collision} describes the change in the solid angle between ${\bf q}$ and ${\bf q}'$.

Solving the full collision integral is a complicated task. Here, we use an approximate method to study small oscillations around the equilibrium distribution (see, e.g., Refs.~\cite{chiacchiera09,pantel12} for more details). To this end, we expand the distribution function as
\begin{align}
&n_\sigma({\bf r}, {\bf p}, t) = n^{\rm eq}_{\sigma}({\bf r}, {\bf p}) \nonumber \\
&\qquad + n^{\rm eq}_{\sigma}({\bf r}, {\bf p}) (1-n^{\rm eq}_{\sigma}({\bf r}, {\bf p})) \Phi_\sigma({\bf r}, {\bf p},t) , \label{eq:smallpert}
\end{align}
where $n^{\rm eq}_{\sigma}({\bf r}, {\bf p})$ is the equilibrium distribution~\eqref{eq:occ}. The prefactor in Eq.~\eqref{eq:smallpert} is chosen such that $\Phi_\sigma$ can be interpreted as a potential perturbation that corrects the quasiparticle energy $\varepsilon_\sigma$. The collision integral then reads:
\begin{align}
 &I_\sigma[\Phi_\sigma, \Phi_{\bar{\sigma}}] = \int \frac{d^3p_1}{(2\pi)^3} \, \int d\Omega \, \frac{d\sigma}{d\Omega} \, |{\bf v}_{\rm rel}| \nonumber \\
 &\times n^{\rm eq}_\sigma({\bf r}, {\bf p}) \, n^{\rm eq}_{\bar{\sigma}}({\bf r}, {\bf p}_1) \, (1 - n^{\rm eq}_\sigma({\bf r}, {\bf p}')) \, (1 - n^{\rm eq}_{\bar{\sigma}}({\bf r}, {\bf p}_1')) \nonumber \\
&\times \left[\Phi_\sigma({\bf r},{\bf p},t) + \Phi_{\bar{\sigma}}({\bf r},{\bf p}_1,t) - \Phi_\sigma({\bf r},{\bf p}',t) - \Phi_{\bar{\sigma}}({\bf r},{\bf p}_1',t)\right] .
\end{align}
This complicated kinetic equation can be solved approximately by expanding the perturbation $\Phi_\sigma({\bf r}, {\bf p})$ in a suitably chosen set of basis functions
\begin{align}
\Phi_\sigma({\bf r}, {\bf p}, t) = e^{-i\omega t} \sum_j c_{j,\sigma} \psi_{j,\sigma}({\bf r}, {\bf p}) ,
\end{align}
where we assume a harmonic time-dependence with frequency $\omega$. Substituting this form in Eq.~\eqref{eq:kinetic}, multiplying by $\psi_i({\bf r}, {\bf p})$, and integrating over ${\bf r}$ and ${\bf p}$ reduces the kinetic equation to a set of linear equations of the form
\begin{align}
\omega M_1 + M_2 + M_3 + C &= 0 ,
\end{align}
where $M_1, M_2,$ and $C$ are matrices with coefficients
\begin{align}
(M_1)^{\sigma\sigma'}_{ij} &= - i \int \frac{d^3r d^3p}{(2\pi)^3} \, \psi_{i,\sigma}({\bf r}, {\bf p}) \, n^{\rm eq}_{\sigma'}({\bf r}, {\bf p}) \nonumber \\
&\quad \times (1-n^{\rm eq}_{\sigma'}({\bf r}, {\bf p})) \psi_{j,\sigma'}({\bf r}, {\bf p})\\
(M_2)^{\sigma\sigma'}_{ij} &=  \int \frac{d^3r d^3p}{(2\pi)^3} \, \psi_{i,\sigma}({\bf r}, {\bf p}) \, \Bigl[(\nabla_{\bf p} \varepsilon_\sigma({\bf r}, {\bf p})) \cdot \nabla_{\bf r} \nonumber \\
&\quad- (\nabla_{\bf r} \varepsilon_\sigma({\bf r}, {\bf p})) \cdot \nabla_{\bf p}\bigr] n^{\rm eq}_{\sigma'}({\bf r}, {\bf p}) \nonumber \\
&\quad \times (1-n^{\rm eq}_{\sigma'}({\bf r}, {\bf p})) \psi_{j,\sigma'}({\bf r}, {\bf p}) \\
(M_3)^{\sigma\sigma'}_{ij} &= - \int \frac{d^3r d^3p}{(2\pi)^3} \, \psi_{i,\sigma}({\bf r}, {\bf p}) \, (\nabla_{\bf p} n^{\rm eq}_{\sigma}({\bf r}, {\bf p})) \cdot \nonumber \\
&\quad \times \biggl[\nabla_{\bf r} \frac{\partial U_{eq,\sigma}}{\partial n^{\rm eq}_{\sigma'}({\bf r})} \delta n_{j,\sigma'}({\bf r})\biggr] \\
(C)^{\sigma\sigma'}_{ij} &= \int \frac{d^3r d^3p}{(2\pi)^3} \, \psi_{i,\sigma}({\bf r}, {\bf p}) \, I_\sigma[\psi_{j,\sigma'}({\bf r}, {\bf p}),0] ,
\end{align}
where 
\begin{align}
\delta n_{j,\sigma'}({\bf r}) &= \int \frac{d^3p}{(2\pi)^3} n^{\rm eq}_{\sigma'}({\bf r}, {\bf p}) (1-n^{\rm eq}_{\sigma'}({\bf r}, {\bf p})) \psi_{j,\sigma'}({\bf r}, {\bf p}) .
\end{align}
The eigenmodes $\omega$ are obtained by computing the matrices $M_1, M_2, M_3,$ and $C$ numerically and solving the eigenvalue problem for the matrix $- M_1^{-1} [M_2 + M_3 + C]$. For the breathing mode oscillation, a suitable set of basis functions is
\begin{align}
\psi_{1,\sigma} &= x^2 + y^2 \\
\psi_{2,\sigma} &= z^2 \\
\psi_{3,\sigma} &= x p_x + y p_y \\
\psi_{4,\sigma} &= z p_z \\
\psi_{5,\sigma} &= p_x^2 + p_y^2 \\
\psi_{6,\sigma} &= p_z^2 .
\end{align}
The computation of the moments is intricate, and we relegate the details of this calculation and the results to App.~\ref{sec:moments}.

\subsection{In-phase mode}

We first discuss the lowest-frequency breathing mode. For a weakly imbalanced Fermi gas, this mode corresponds to an in-phase breathing mode of both spin species, which changes at large polarization, where it describes the breathing mode of the majority species. In the following, we choose the same parameters as in the experiment~\cite{nascimbene09}: the aspect ratio of the trap is $\lambda = \omega_z/\omega_r = 0.075$ and we explore the unitary limit $a\to \infty$ at low temperatures. The temperature scale is set by the majority density as $T_F = \omega_0 (6N_\uparrow)^{1/3}$ with $\omega_0 = (\omega_z \omega_r^2)^{1/3}$, which corresponds to the Fermi energy of a noninteracting trapped single-component gas. 

\begin{figure}[t]
\subfigure[]{
\scalebox{0.75}{\includegraphics{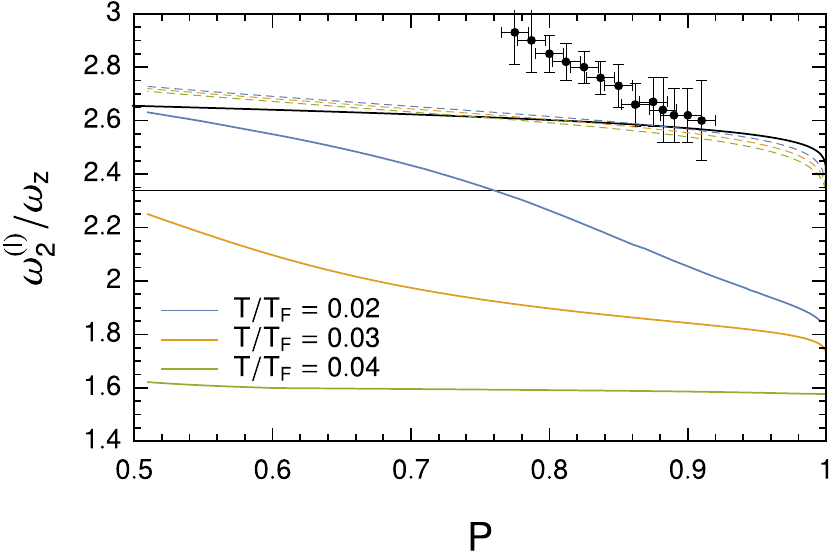}}\qquad\quad}\\
\subfigure[]{
\scalebox{0.75}{\includegraphics{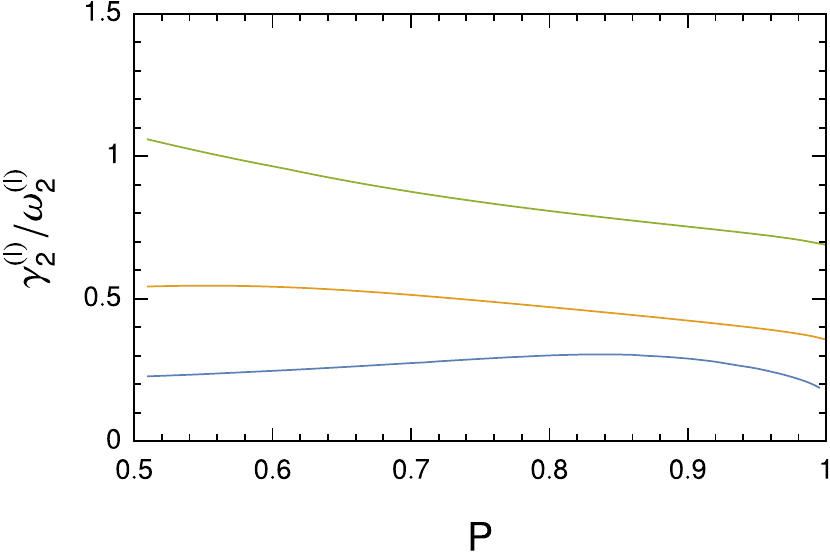}}\qquad\quad}
\caption{(a) Frequency and (b) damping of the longitudinal out-of-phase breathing mode of an imbalanced Fermi gas in an anisotropic trap with aspect ratio $\lambda=0.075$ as a function of polarization for three different temperatures $T/T_F = 0.02,0.03$ and $0.04$. For comparison, we show the collisionless results as dashed lines. The continuous black line indicates the zero temperature scaling result of Ref.~\cite{recati12} and the thin black line denotes the collisionless limit Eq.~\eqref{eq:collfreq}.  The experimental data points are taken from Ref.~\cite{nascimbene09}.}
\label{fig:2}
\end{figure}

Figure~\ref{fig:1} shows the in-phase breathing mode as a function of polarization for several temperatures $T/T_F = 0.06,0.08,0.1,0.12,0.15$, and $0.18$. Figure~\ref{fig:1}(a) shows the collective mode frequency, which clearly displays a crossover between a hydrodynamic and a collisionless limit. The oscillation frequency in the collisionless limit $P\to 1$ is equal to twice the trap frequency, $\omega_{\rm cl} = 2 \omega_z$. In the hydrodynamic limit $P\ll 1$, the frequency can be estimated by taking moments of the $z^2$ dynamic structure factor yielding $\omega_{\rm hd} = \sqrt{12/5} \omega_z$~\cite{nascimbene09}. Both limiting cases are indicated by thin black lines. The solution of the Fermi liquid kinetic theory is in good agreement with the experimental measurements and provides an accurate description of the collisionless-to-hydrodynamic crossover with the experimental parameters. At unitarity, the Clogston-Chandrasekar limit is at $P=0.75$, which puts a lower limit on the applicability of our theory. Nevertheless, even below that,  in the superfluid phase, there is only a small quantitative discrepancy with the experiment. Figure~\ref{fig:1}(b) shows the damping of the collective mode frequency. Again, our theoretical calculations are in good quantitative agreement with the experiment~\cite{nascimbene09}, with optimal agreement at a temperature $T \approx 0.12 T_F$, the same optimal temperature as for the collective mode frequency.

Finally, in Fig.~\ref{fig:1}(c), we show a reduced plot of damping versus frequency, which does not contain the polarization. All results approximately collapse onto a single scaling curve. This indicates the presence of a single dominant relaxation time $\tau$ and is consistent with a thermodynamic argument for the crossover, which predicts that frequency and damping satisfy~\cite{*[{}] [{, \S78.}] landau66}
\begin{align}
 \omega^2 = \omega_{\rm cl}^2 + \frac{\omega_{\rm hd}^2 - \omega_{\rm cl}^2}{1 + i \omega \tau} , \label{eq:intform}
\end{align}
where $\omega$ is a complex number, the real part of which sets the mode frequency $\omega_1$ and the imaginary part sets the damping $\gamma_1$. This scaling solution is shown as a black dashed line in Fig.~\ref{fig:1}(c) for comparison.

In addition to the longitudinal in-phase breathing mode, there is also a radial in-phase oscillation. Because the radial trapping frequency is much larger than the longitudinal frequency, $\omega_r \gg \omega_z$, collisions are much less efficient here (as can be seen, for example, from Eq.~\eqref{eq:intform}). For all temperatures in our calculation, the oscillation is only very weakly damped and remains close to the collisionless value $\omega_1^{(r)} = 2 \omega_r$ for all polarizations. We do not plot this mode.

\begin{figure}[t!]
\subfigure[]{\scalebox{0.75}{\includegraphics{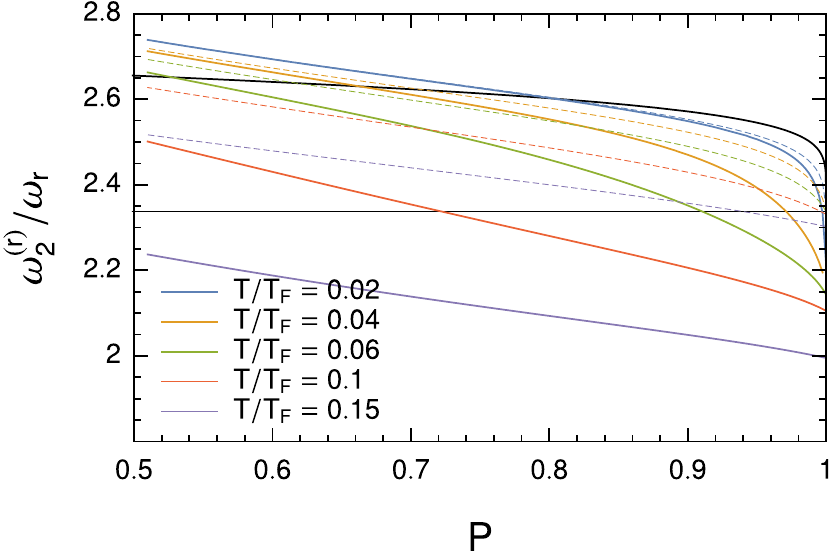}}\qquad\quad}\\
\subfigure[]{\scalebox{0.75}{\includegraphics{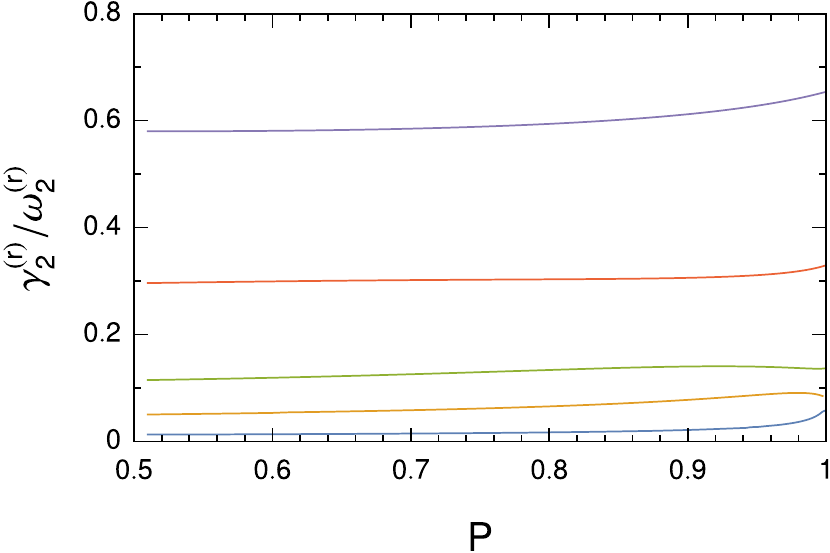}}\qquad\quad}
\caption{(a) Frequency and (b) damping of the radial out-of-phase breathing mode of an imbalanced Fermi gas in an anisotropic trap with aspect ratio $\lambda=0.075$ as a function of polarization for different temperatures $T/T_F = 0.02,0.04,0.06,0.1$ and $0.15$. For comparison, we show the collisionless results as dashed lines. The continuous black line indicates the zero temperature scaling result of Ref.~\cite{recati12} and the thin black line denotes the collisionless limit Eq.~\eqref{eq:collfreq}.}
\label{fig:3}
\end{figure}

\subsection{Out-of-phase mode}

There is a second higher-frequency breathing mode excitation for each trap direction, which corresponds to an out-of-phase breathing mode at small polarization and reduces to an oscillation of the minority atoms at large polarization. This limit is of particular interest as the collisionless oscillation frequency, Eq.~\eqref{eq:collfreq}, depends on the polaron mass.

Figure~\ref{fig:2} shows our results for the frequency and damping of the longitudinal out-of-phase for three different temperatures $T/T_F = 0.02, 0.03$, and $0.04$. Different from the in-phase oscillation, the collisionless limit cannot be reached by changing the polarization, and we find that the mode is very strongly damped at any polarization. Indeed, for any larger temperatures $T/T_F > 0.05$, the mode is completely overdamped. For comparision, we include the collisionless frequencies as dashed lines in Fig.~\ref{fig:2}(a). We find that at small temperature and high polarization, the difference of the breathing mode from the single polaron frequency [Eq.~\eqref{eq:collfreq}] is proportional to the radius of the minority cloud, which depends on the polarization as
\begin{equation}
\frac{R_\downarrow}{R_\uparrow} \sim \left(\frac{1-P}{1+P} \right)^{1/6} .
\end{equation}
This result was previously established by Recati and Stringari~\cite{recati10}, who analyzed the mode at zero temperature neglecting collisions by combining a scaling ansatz and a density functional for the ground state energy of the imbalanced gas. We show their result in Fig.~\ref{fig:2} for comparison (black line). At finite temperature, this effect is less pronounced and decreases with increasing temperature, and our calculation suggests a linear dependence of the collisionless breathing mode frequency on the polarization for $P < 0.9$.
 
The calculated frequencies are at odds with the experimental measurements~\cite{nascimbene09} (black points in Fig.~\ref{fig:2}). While already the collisionless results differs from the experimental data, it was suggested in~\cite{recati10} that collisions could be responsible for this discrepancy. Our calculations, which do include collisions, would seem to refute this claim. We find that the damping of this mode is very significant, to the extent that it would be overdamped for all values of $P$ at the experimental temperatures, rendering it difficult to observe, in contradiction with the experiment.

The radial out-of-phase breathing mode persists over a larger range of temperatures since collisions are less efficient compared to the longitudinal oscillation. Figure~\ref{fig:3} shows the frequency and damping of this mode for temperatures $T/T_F = 0.02,0.04,0.06,0.1$, and $0.15$. Collisions decrease the oscillation frequency compared to the collisionless case (dashed lines) with a strong damping at any polarization. The oscillation reduces to the collisionless frequency only at small temperatures.

\section{Summary and Conclusions}

In conclusion, we have studied the collective breathing modes of a strongly imbalanced unitary Fermi gas, assuming that it can be described as an interacting gas of minority polarons and majority atoms. We have solved the kinetic equation in an elongated harmonic trap taking into account quasiparticle collisions. For the in-phase breathing mode, our results provide an accurate description of both frequency and damping observed in the experiment by Nascimb\`ene {\it et al.}~\cite{nascimbene09}. The theory displays a crossover between a collisionless limit at large polarization, where the mode frequency $\omega$ is much larger than the inverse collision time $1/\tau$, $\omega \tau \gg 1$, and a hydrodynamic limit $\omega \tau \ll 1$, where single excitations decay rapidly. Our theory appears to be reliable down to the critical polarization $P\sim 0.7$, below which a superfluid core forms at the trap center. By contrast, our results for the out-of-phase breathing mode oscillation differ from the findings in~\cite{nascimbene09}. While our results are consistent with predictions from a scaling ansatz for the collisionless gas~\cite{recati12}, taking into account collisions does not resolve the discrepancy between theory and experiment.

\begin{acknowledgments}
J.H. is supported by Gonville and Caius College, Cambridge. O.G. is supported by the National Science Foundation (Grant PHY-1314735) and the US-Israel Binational Science Foundation (Grants 2014262 and 2016087).
\end{acknowledgments}

\appendix
\section{Method of moments}\label{sec:moments}

We determine the lowest breathing mode excitations of a spin-imbalanced Fermi gas by solving the linearized Boltzmann equation using the method of moments. This appendix describes the details of the calculation.

We define the potential energy per spin species:
\begin{align}
E_{\rm pot, \sigma} &= \int d^3r \, V_{\rm trap} \, n^{\rm eq}_\sigma 
\end{align}
and the kinetic energy
\begin{align}
 E_{\rm kin, \sigma} &= \int \frac{d^3r d^3p}{(2\pi)^3} \, \frac{p^2}{2 m_\sigma} \, n^{\rm eq}_\sigma .
\end{align}
They are related through the virial theorem
\begin{align}
 \frac{E_{{\rm kin},\sigma}}{E_{{\rm pot},\sigma}} &= 1 - \tilde{\chi}_\sigma ,
\end{align}
where $\chi_\sigma$ is defined as
\begin{align}
\tilde{\chi}_\sigma &= - \frac{1}{2 E_{\rm pot, \sigma}} \int d^3\tilde{r} \, n^{\rm eq}_\sigma \tilde{r} \, \frac{\partial U_\sigma^{\rm eq}}{\partial \tilde{r}} .
\end{align}
The eigenmodes are determined by solving the equation
\begin{align}
 \det (A+B) &= 0 .
\end{align}
$A$ is the matrix of moments of the streaming term, and $B$ the matrix for the collision integral. They are:

\begin{widetext}
 \makeatletter
    \def\tagform@#1{\maketag@@@{\normalsize(#1)\@@italiccorr}}
\makeatother
\tiny
\begin{align}
A^{\sigma\sigma} &= \left( \begin{matrix}
\dfrac{2 i \omega \omega_0 (1 + \varphi_\sigma)}{\omega_r^2} &
\dfrac{i \omega \omega_0 (1 + \varphi_\sigma)}{2 \omega_r^2} &
1 &
0 &
\dfrac{i \omega}{\omega_0} \dfrac{m_\sigma}{m} &
\dfrac{i \omega}{2 \omega_0} \dfrac{m_\sigma}{m} \\[2ex]
\dfrac{2 i \omega \omega_0 (1 + \varphi_\sigma)}{\omega_r^2} &
\dfrac{3 i \omega \omega_0 (1 + \varphi_\sigma)}{\omega_r^2} &
0 &
2 &
2 \dfrac{i \omega}{\omega_0} \dfrac{m_\sigma}{m} &
\dfrac{i \omega}{\omega_0} \dfrac{m_\sigma}{m} \\[2ex]
2 (1 + 2 \varphi_{1,\sigma\sigma} - \varphi_{3,\sigma\sigma}) &
\varphi_{1,\sigma\sigma} - \varphi_{3,\sigma\sigma} &
 - \dfrac{i \omega}{\omega_0} \dfrac{m_\sigma}{m} &
0 &
- 2 \dfrac{\omega_r^2}{\omega_0^2} \dfrac{m_\sigma}{m} \left(1 - \tilde{\chi}_\sigma - \chi_{\sigma\sigma} + 2 \chi_{\sigma\sigma}'\right) &
\dfrac{\omega_r^2}{\omega_0^2} \dfrac{m_{\bar{\sigma}}}{m} \left(\chi_{\sigma\sigma} - 2 \chi_{\sigma\sigma}'\right) \\[2ex]
2 (\varphi_{1,\sigma\sigma} - \varphi_{3,\sigma\sigma}) &
2 + 3 \varphi_{1,\sigma\sigma} - \varphi_{3,\sigma\sigma} &
0 &
- \dfrac{i \omega}{\omega_0 \lambda^2} \dfrac{m_\sigma}{m} &
2 \dfrac{\omega_r^2}{\omega_0^2} \dfrac{m_\sigma}{m} \left(\chi_{\sigma\sigma} - 2 \chi_{\sigma\sigma}'\right) &
- 2 \dfrac{\omega_r^2}{\omega_0^2} \dfrac{m_\sigma}{m} \left(1 - \tilde{\chi}_\sigma - \chi_{\sigma\sigma} + 2 \chi_{\sigma\sigma}'\right) \\[2ex]
\dfrac{i m \omega \omega_0}{m_\sigma \omega_r^2 (1 - \tilde{\chi}_\sigma)} &
\dfrac{i m \omega \omega_0}{2 m_\sigma \omega_r^2 (1 - \tilde{\chi}_\sigma)} &
- \dfrac{m}{m_\sigma} &
0 &
2 \dfrac{i \omega}{\omega_0} &
\dfrac{i \omega }{2 \omega_0} \\[2ex]
\dfrac{i m \omega \omega_0}{m_\sigma \omega_r^2 (1 - \tilde{\chi}_\sigma)} &
\dfrac{i m \omega \omega_0}{2 m_\sigma \omega_r^2 (1 - \tilde{\chi}_\sigma)} &
0 &
- \dfrac{m}{m_\sigma} &
\dfrac{i \omega}{\omega_0} &
\dfrac{3}{2} \dfrac{i \omega}{\omega_0} \end{matrix} \right) , \\[5ex]
A^{\sigma\bar{\sigma}} &= \left( \begin{matrix}
0 &
0 &
0 &
0 &
0 &
0 \\[2ex]
0 &
0 &
0 &
0 &
0 &
0 \\[2ex]
2 (2 \varphi_{1,\sigma\bar{\sigma}} - \varphi_{3,\sigma\bar{\sigma}}) &
\varphi_{1,\sigma\bar{\sigma}} - \varphi_{3,\sigma\bar{\sigma}} &
0 &
0 &
2 \dfrac{\omega_r^2}{\omega_0^2} \dfrac{m_{\bar{\sigma}}}{m} (\chi_{\sigma\bar{\sigma}} - 2 \chi_{\sigma\bar{\sigma}}') &
\dfrac{\omega_r^2}{\omega_0^2} \dfrac{m_{\bar{\sigma}}}{m} (\chi_{\sigma\bar{\sigma}} - 2 \chi_{\sigma\bar{\sigma}}') \\[2ex]
2 (\varphi_{1,\sigma\bar{\sigma}} - \varphi_{3,\sigma\bar{\sigma}}) &
3 \varphi_{1,\sigma\bar{\sigma}} - \varphi_{3,\sigma\bar{\sigma}} &
0 &
0 &
2 \dfrac{\omega_r^2}{\omega_0^2} \dfrac{m_{\bar{\sigma}}}{m} (\chi_{\sigma\bar{\sigma}} - 2 \chi_{\sigma\bar{\sigma}}') &
\dfrac{\omega_r^2}{\omega_0^2} \dfrac{m_{\bar{\sigma}}}{m} (\chi_{\sigma\bar{\sigma}} - 2 \chi_{\sigma\bar{\sigma}}') \\[2ex]
0 &
0 &
0 &
0 &
0 &
0 \\[2ex]
0 &
0 &
0 &
0 &
0 &
0 \end{matrix} \right) , \\[5ex]
B^{\sigma\tau} &=\begin{pmatrix}0&0&0&0&0&0\\0&0&0&0&0&0\\0&0&
\dfrac{\pm 1}{\tau_{A\sigma}}
&0&0&0\\0&0&0&
\dfrac{\pm 1}{\lambda^2 \tau_{A\sigma}}
&0&0\\0&0&0&0&
\dfrac{-1}{\tau_{B\sigma}} - \dfrac{m_\sigma \pm m_\tau}{M} \dfrac{1}{\tau_{C\sigma}} \mp \dfrac{m_\sigma m_\tau}{M^2} \left(\dfrac{1}{\tau_{D\sigma}} + \dfrac{1}{\tau_{E\sigma}}\right)
&
\dfrac{1}{\tau_{B\sigma}} + \dfrac{m_\sigma \pm m_\tau}{M} \dfrac{1}{\tau_{C\sigma}} \pm \dfrac{m_\sigma m_\tau}{M^2} \left(\dfrac{1}{\tau_{D\sigma}} - \dfrac{1}{\tau_{E\sigma}}\right)
\\[2ex]
0&0&0&0&
\dfrac{2}{\tau_{B\sigma}} + \dfrac{m_\sigma \pm m_\tau}{M} \dfrac{2}{\tau_{C\sigma}} \pm \dfrac{2 m_\sigma m_\tau}{M^2} \left(\dfrac{1}{\tau_{D\sigma}} - \dfrac{1}{\tau_{E\sigma}}\right)
&
- \dfrac{2}{\tau_{B\sigma}} - \dfrac{m_\sigma \pm m_\tau}{M} \dfrac{2}{\tau_{C\sigma}} \mp \dfrac{m_\sigma m_\tau}{M^2} \dfrac{2}{\tau_{D\sigma}}\end{pmatrix},
\end{align}
\normalsize
where the upper sign for $B$ applies if $\sigma =\tau$ and the lower if $\sigma\neq\tau$ and we define the dimensionless quantities (use rescaled coordinates $\tilde{r}_i = \omega_i r_i/\omega_0$):
\begin{align}
\chi_{\sigma\tau} &= - \frac{1}{2 E_{\rm pot, \sigma}} \int d^3\tilde{r} \, n^{\rm eq}_\tau \, \tilde{r} \, \frac{\partial U_\sigma^{\rm eq}}{\partial n^{\rm eq}_\tau} \, \frac{\partial n^{\rm eq}_\sigma}{\partial \tilde{r}} \\ 
\chi_{\sigma\tau}' &= \frac{3}{4 E_{\rm pot, \sigma}} \int d^3\tilde{r} \, n^{\rm eq}_\sigma \, n^{\rm eq}_\tau \, \frac{\partial U_\sigma^{\rm eq}}{\partial n^{\rm eq}_\tau} \label{eq:momenta} \\
\varphi_{\sigma} &= \frac{1}{10 E_{\rm pot, \sigma}} \int d^3\tilde{r} \, \tilde{r}^2 \frac{\partial n^{\rm eq}_\sigma}{\partial \tilde{r}} \, \frac{\partial U_\sigma^{\rm eq}}{\partial \tilde{r}} \, \frac{1}{1 + \dfrac{1}{m \omega_0^2 \tilde{r}} \dfrac{\partial U_\sigma^{\rm eq}}{\partial \tilde{r}}} \\
\varphi_{1,\sigma\tau} &= \frac{1}{10 E_{\rm pot, \sigma}} \int d^3\tilde{r} \, \tilde{r}^2 \frac{\partial n^{\rm eq}_\sigma}{\partial \tilde{r}} \, \frac{\partial n^{\rm eq}_\tau}{\partial \tilde{r}}\,\frac{\partial U_\sigma^{\rm eq}}{\partial n^{\rm eq}_\tau} \, \frac{1}{1 + \dfrac{1}{m \omega_0^2 \tilde{r}} \dfrac{\partial U_\tau^{\rm eq}}{\partial \tilde{r}}} \\
\varphi_{3,\sigma\tau} &= - \frac{1}{2 E_{\rm pot, \sigma}} \int d^3\tilde{r} \, \tilde{r} n^{\rm eq}_\sigma \, \frac{\partial n^{\rm eq}_\tau}{\partial \tilde{r}} \, \frac{\partial U_\sigma^{\rm eq}}{\partial n^{\rm eq}_\tau} \, \frac{1}{1 + \dfrac{1}{m \omega_0^2 \tilde{r}} \dfrac{\partial U_\tau^{\rm eq}}{\partial \tilde{r}}} . \label{eq:momentb}
\end{align}
The various relaxation times can be calculated along the lines of Ref.~\cite{vichi00}:
\begin{align}
\frac{1}{\tau_{i\sigma}} &= \int_0^\infty d\tilde{r} \, \tilde{r}^2 \, \int_0^\infty dP \, P^2 \, \int_0^\infty dq \, q^5 \, \frac{d\sigma}{d\Omega} \, \int_{-1}^1 d(x,y) \, n_\downarrow({\bf r}, {\bf p}) \, n_\uparrow({\bf r}, {\bf p}_1) \, (1 - n_\downarrow({\bf r}, {\bf p}')) \, (1 - n_\uparrow({\bf r}, {\bf p}_1')) \times g_{i,\sigma}(x,y) ,
\end{align}
where $g_{A,\sigma}(x,y) = \frac{\beta \omega_0 \tilde{r}^2 (1 - x y)}{6 \pi^2 m_{\rm red} E_{{\rm pot},\sigma}}$ and
\begin{align}
g_{i,\sigma}(x,y) &= \frac{\beta}{\pi^2 m_{\rm red} m_\sigma^2 \omega_0 E_{{\rm kin},\sigma}}
\begin{cases}
\dfrac{q^2 }{40} (1+x^2+y^2-3x^2y^2) & i=B \\[2ex]
\dfrac{q P}{30} (x+y) (1 - x y) & i=C \\[2ex]
\dfrac{P^2}{10} (1+x^2+y^2-3xy) & i=D \\[2ex]
\dfrac{P^2}{4} (x-y)^2 & i=E
\end{cases} .
\end{align}
\end{widetext}

\bibliography{bib}

\begin{thebibliography}{35}%
\makeatletter
\providecommand \@ifxundefined [1]{%
 \@ifx{#1\undefined}
}%
\providecommand \@ifnum [1]{%
 \ifnum #1\expandafter \@firstoftwo
 \else \expandafter \@secondoftwo
 \fi
}%
\providecommand \@ifx [1]{%
 \ifx #1\expandafter \@firstoftwo
 \else \expandafter \@secondoftwo
 \fi
}%
\providecommand \natexlab [1]{#1}%
\providecommand \enquote  [1]{``#1''}%
\providecommand \bibnamefont  [1]{#1}%
\providecommand \bibfnamefont [1]{#1}%
\providecommand \citenamefont [1]{#1}%
\providecommand \href@noop [0]{\@secondoftwo}%
\providecommand \href [0]{\begingroup \@sanitize@url \@href}%
\providecommand \@href[1]{\@@startlink{#1}\@@href}%
\providecommand \@@href[1]{\endgroup#1\@@endlink}%
\providecommand \@sanitize@url [0]{\catcode `\\12\catcode `\$12\catcode
  `\&12\catcode `\#12\catcode `\^12\catcode `\_12\catcode `\%12\relax}%
\providecommand \@@startlink[1]{}%
\providecommand \@@endlink[0]{}%
\providecommand \url  [0]{\begingroup\@sanitize@url \@url }%
\providecommand \@url [1]{\endgroup\@href {#1}{\urlprefix }}%
\providecommand \urlprefix  [0]{URL }%
\providecommand \Eprint [0]{\href }%
\providecommand \doibase [0]{http://dx.doi.org/}%
\providecommand \selectlanguage [0]{\@gobble}%
\providecommand \bibinfo  [0]{\@secondoftwo}%
\providecommand \bibfield  [0]{\@secondoftwo}%
\providecommand \translation [1]{[#1]}%
\providecommand \BibitemOpen [0]{}%
\providecommand \bibitemStop [0]{}%
\providecommand \bibitemNoStop [0]{.\EOS\space}%
\providecommand \EOS [0]{\spacefactor3000\relax}%
\providecommand \BibitemShut  [1]{\csname bibitem#1\endcsname}%
\let\auto@bib@innerbib\@empty
\bibitem [{\citenamefont {Landau}(1957{\natexlab{a}})}]{landau57a}%
  \BibitemOpen
  \bibfield  {author} {\bibinfo {author} {\bibfnamefont {L.~D.}\ \bibnamefont
  {Landau}},\ }\bibfield  {title} {\enquote {\bibinfo {title} {{Theory of the
  Fermi Liquid}},}\ }\href@noop {} {\bibfield  {journal} {\bibinfo  {journal}
  {Sov. Phys. JETP}\ }\textbf {\bibinfo {volume} {3}},\ \bibinfo {pages} {920}
  (\bibinfo {year} {1957}{\natexlab{a}})}\BibitemShut {NoStop}%
\bibitem [{\citenamefont {Landau}(1957{\natexlab{b}})}]{landau57b}%
  \BibitemOpen
  \bibfield  {author} {\bibinfo {author} {\bibfnamefont {L.~D.}\ \bibnamefont
  {Landau}},\ }\bibfield  {title} {\enquote {\bibinfo {title} {{Oscillations in
  a Fermi Liquid}},}\ }\href@noop {} {\bibfield  {journal} {\bibinfo  {journal}
  {Sov. Phys. JETP}\ }\textbf {\bibinfo {volume} {5}},\ \bibinfo {pages} {101}
  (\bibinfo {year} {1957}{\natexlab{b}})}\BibitemShut {NoStop}%
\bibitem [{\citenamefont {Landau}(1959)}]{landau59}%
  \BibitemOpen
  \bibfield  {author} {\bibinfo {author} {\bibfnamefont {L.~D.}\ \bibnamefont
  {Landau}},\ }\bibfield  {title} {\enquote {\bibinfo {title} {{On the Theory
  of the Fermi Liquid}},}\ }\href@noop {} {\bibfield  {journal} {\bibinfo
  {journal} {Sov. Phys. JETP}\ }\textbf {\bibinfo {volume} {35}},\ \bibinfo
  {pages} {70} (\bibinfo {year} {1959})}\BibitemShut {NoStop}%
\bibitem [{\citenamefont {Pines}\ and\ \citenamefont
  {Nozi\`eres}(1994)}]{pines94}%
  \BibitemOpen
  \bibfield  {author} {\bibinfo {author} {\bibfnamefont {D.}~\bibnamefont
  {Pines}}\ and\ \bibinfo {author} {\bibfnamefont {P.}~\bibnamefont
  {Nozi\`eres}},\ }\href@noop {} {\emph {\bibinfo {title} {Theory Of Quantum
  Liquids, Volume I: Normal Fermi Liquids}}}\ (\bibinfo  {publisher} {Westview
  Press},\ \bibinfo {year} {1994})\BibitemShut {NoStop}%
\bibitem [{\citenamefont {Lifshitz}\ and\ \citenamefont
  {Pitaevskii}(2006)}]{landau06}%
  \BibitemOpen
  \bibfield  {author} {\bibinfo {author} {\bibfnamefont {E.~M.}\ \bibnamefont
  {Lifshitz}}\ and\ \bibinfo {author} {\bibfnamefont {L.~P.}\ \bibnamefont
  {Pitaevskii}},\ }\href@noop {} {\emph {\bibinfo {title} {{Course of
  Theoretical Physics, Vol. 9: Statistical Physics, Part 2}}}}\ (\bibinfo
  {publisher} {Butterworth-Heinemann},\ \bibinfo {year} {2006})\BibitemShut
  {NoStop}%
\bibitem [{\citenamefont {Giuliani}\ and\ \citenamefont
  {Vignale}(2005)}]{giuliani05}%
  \BibitemOpen
  \bibfield  {author} {\bibinfo {author} {\bibfnamefont {G.}~\bibnamefont
  {Giuliani}}\ and\ \bibinfo {author} {\bibfnamefont {G.}~\bibnamefont
  {Vignale}},\ }\href@noop {} {\emph {\bibinfo {title} {{Quantum Theory of the
  Electron Liquid}}}}\ (\bibinfo  {publisher} {Cambridge University Press},\
  \bibinfo {year} {2005})\BibitemShut {NoStop}%
\bibitem [{\citenamefont {Lobo}\ \emph {et~al.}(2006)\citenamefont {Lobo},
  \citenamefont {Recati}, \citenamefont {Giorgini},\ and\ \citenamefont
  {Stringari}}]{lobo06}%
  \BibitemOpen
  \bibfield  {author} {\bibinfo {author} {\bibfnamefont {C.}~\bibnamefont
  {Lobo}}, \bibinfo {author} {\bibfnamefont {A.}~\bibnamefont {Recati}},
  \bibinfo {author} {\bibfnamefont {S.}~\bibnamefont {Giorgini}}, \ and\
  \bibinfo {author} {\bibfnamefont {S.}~\bibnamefont {Stringari}},\ }\bibfield
  {title} {\enquote {\bibinfo {title} {{Normal State of a Polarized Fermi Gas
  at Unitarity}},}\ }\href {\doibase 10.1103/PhysRevLett.97.200403} {\bibfield
  {journal} {\bibinfo  {journal} {Phys. Rev. Lett.}\ }\textbf {\bibinfo
  {volume} {97}},\ \bibinfo {pages} {200403} (\bibinfo {year}
  {2006})}\BibitemShut {NoStop}%
\bibitem [{\citenamefont {Zwerger}(2016)}]{zwerger16}%
  \BibitemOpen
  \bibfield  {author} {\bibinfo {author} {\bibfnamefont {W.}~\bibnamefont
  {Zwerger}},\ }\bibfield  {title} {\enquote {\bibinfo {title} {{Strongly
  Interacting Fermi Gases}},}\ }in\ \href@noop {} {\emph {\bibinfo {booktitle}
  {Proceedings of the International School of Physics ``Enrico Fermi" - Course
  191 ``Quantum Matter at Ultralow Temperatures"}}},\ \bibinfo {editor} {edited
  by\ \bibinfo {editor} {\bibfnamefont {M.}~\bibnamefont {Inguscio}}, \bibinfo
  {editor} {\bibfnamefont {W.}~\bibnamefont {Ketterle}}, \bibinfo {editor}
  {\bibfnamefont {S.}~\bibnamefont {Stringari}}, \ and\ \bibinfo {editor}
  {\bibfnamefont {G.}~\bibnamefont {Roati}}}\ (\bibinfo  {publisher} {IOS
  Press, Amsterdam; SIF Bologna},\ \bibinfo {year} {2016})\BibitemShut
  {NoStop}%
\bibitem [{\citenamefont {Chevy}\ and\ \citenamefont
  {Salomon}(2012)}]{chevy12}%
  \BibitemOpen
  \bibfield  {author} {\bibinfo {author} {\bibfnamefont {F.}~\bibnamefont
  {Chevy}}\ and\ \bibinfo {author} {\bibfnamefont {C.}~\bibnamefont
  {Salomon}},\ }\bibfield  {title} {\enquote {\bibinfo {title} {{Thermodynamics
  of Fermi Gases}},}\ }in\ \href@noop {} {\emph {\bibinfo {booktitle} {The
  BCS--BEC Crossover and the Unitary Fermi Gas}}},\ \bibinfo {editor} {edited
  by\ \bibinfo {editor} {\bibfnamefont {W.}~\bibnamefont {Zwerger}}}\ (\bibinfo
   {publisher} {Springer},\ \bibinfo {year} {2012})\ Chap.~\bibinfo {chapter}
  {11}\BibitemShut {NoStop}%
\bibitem [{\citenamefont {Recati}\ and\ \citenamefont
  {Stringari}(2012)}]{recati12}%
  \BibitemOpen
  \bibfield  {author} {\bibinfo {author} {\bibfnamefont {A.}~\bibnamefont
  {Recati}}\ and\ \bibinfo {author} {\bibfnamefont {S.}~\bibnamefont
  {Stringari}},\ }\bibfield  {title} {\enquote {\bibinfo {title} {{Normal Phase
  of Polarised Strongly Interacting Fermi Gases}},}\ }in\ \href@noop {} {\emph
  {\bibinfo {booktitle} {The BCS--BEC Crossover and the Unitary Fermi Gas}}},\
  \bibinfo {editor} {edited by\ \bibinfo {editor} {\bibfnamefont
  {W.}~\bibnamefont {Zwerger}}}\ (\bibinfo  {publisher} {Springer},\ \bibinfo
  {year} {2012})\ Chap.~\bibinfo {chapter} {12}\BibitemShut {NoStop}%
\bibitem [{\citenamefont {Chevy}(2006)}]{chevy06}%
  \BibitemOpen
  \bibfield  {author} {\bibinfo {author} {\bibfnamefont {F.}~\bibnamefont
  {Chevy}},\ }\bibfield  {title} {\enquote {\bibinfo {title} {{Universal Phase
  Diagram of a Strongly Interacting Fermi Gas with Unbalanced Spin
  Populations}},}\ }\href {\doibase 10.1103/PhysRevA.74.063628} {\bibfield
  {journal} {\bibinfo  {journal} {Phys. Rev. A}\ }\textbf {\bibinfo {volume}
  {74}},\ \bibinfo {pages} {063628} (\bibinfo {year} {2006})}\BibitemShut
  {NoStop}%
\bibitem [{\citenamefont {Combescot}\ \emph {et~al.}(2007)\citenamefont
  {Combescot}, \citenamefont {Recati}, \citenamefont {Lobo},\ and\
  \citenamefont {Chevy}}]{combescot07}%
  \BibitemOpen
  \bibfield  {author} {\bibinfo {author} {\bibfnamefont {R.}~\bibnamefont
  {Combescot}}, \bibinfo {author} {\bibfnamefont {A.}~\bibnamefont {Recati}},
  \bibinfo {author} {\bibfnamefont {C.}~\bibnamefont {Lobo}}, \ and\ \bibinfo
  {author} {\bibfnamefont {F.}~\bibnamefont {Chevy}},\ }\bibfield  {title}
  {\enquote {\bibinfo {title} {{Normal State of Highly Polarized Fermi Gases:
  Simple Many-Body Approaches}},}\ }\href {\doibase
  10.1103/PhysRevLett.98.180402} {\bibfield  {journal} {\bibinfo  {journal}
  {Phys. Rev. Lett.}\ }\textbf {\bibinfo {volume} {98}},\ \bibinfo {pages}
  {180402} (\bibinfo {year} {2007})}\BibitemShut {NoStop}%
\bibitem [{\citenamefont {Veillette}\ \emph {et~al.}(2008)\citenamefont
  {Veillette}, \citenamefont {Moon}, \citenamefont {Lamacraft}, \citenamefont
  {Radzihovsky}, \citenamefont {Sachdev},\ and\ \citenamefont
  {Sheehy}}]{veillette08}%
  \BibitemOpen
  \bibfield  {author} {\bibinfo {author} {\bibfnamefont {M.}~\bibnamefont
  {Veillette}}, \bibinfo {author} {\bibfnamefont {E.~G.}\ \bibnamefont {Moon}},
  \bibinfo {author} {\bibfnamefont {A.}~\bibnamefont {Lamacraft}}, \bibinfo
  {author} {\bibfnamefont {L.}~\bibnamefont {Radzihovsky}}, \bibinfo {author}
  {\bibfnamefont {S.}~\bibnamefont {Sachdev}}, \ and\ \bibinfo {author}
  {\bibfnamefont {D.~E.}\ \bibnamefont {Sheehy}},\ }\bibfield  {title}
  {\enquote {\bibinfo {title} {{Radio-Frequency Spectroscopy of a Strongly
  Imbalanced Feshbach-Resonant Fermi Gas}},}\ }\href {\doibase
  10.1103/PhysRevA.78.033614} {\bibfield  {journal} {\bibinfo  {journal} {Phys.
  Rev. A}\ }\textbf {\bibinfo {volume} {78}},\ \bibinfo {pages} {033614}
  (\bibinfo {year} {2008})}\BibitemShut {NoStop}%
\bibitem [{\citenamefont {Combescot}\ and\ \citenamefont
  {Giraud}(2008)}]{combescot08}%
  \BibitemOpen
  \bibfield  {author} {\bibinfo {author} {\bibfnamefont {R.}~\bibnamefont
  {Combescot}}\ and\ \bibinfo {author} {\bibfnamefont {S.}~\bibnamefont
  {Giraud}},\ }\bibfield  {title} {\enquote {\bibinfo {title} {{Normal State of
  Highly Polarized Fermi Gases: Full Many-Body Treatment}},}\ }\href {\doibase
  10.1103/PhysRevLett.101.050404} {\bibfield  {journal} {\bibinfo  {journal}
  {Phys. Rev. Lett.}\ }\textbf {\bibinfo {volume} {101}},\ \bibinfo {pages}
  {050404} (\bibinfo {year} {2008})}\BibitemShut {NoStop}%
\bibitem [{\citenamefont {Prokof'ev}\ and\ \citenamefont
  {Svistunov}(2008{\natexlab{a}})}]{prokofev08a}%
  \BibitemOpen
  \bibfield  {author} {\bibinfo {author} {\bibfnamefont {N.}~\bibnamefont
  {Prokof'ev}}\ and\ \bibinfo {author} {\bibfnamefont {B.}~\bibnamefont
  {Svistunov}},\ }\bibfield  {title} {\enquote {\bibinfo {title}
  {{Fermi-Polaron Problem: Diagrammatic Monte Carlo Method for Divergent
  Sign-Alternating Series}},}\ }\href {\doibase 10.1103/PhysRevB.77.020408}
  {\bibfield  {journal} {\bibinfo  {journal} {Phys. Rev. B}\ }\textbf {\bibinfo
  {volume} {77}},\ \bibinfo {pages} {020408} (\bibinfo {year}
  {2008}{\natexlab{a}})}\BibitemShut {NoStop}%
\bibitem [{\citenamefont {Prokof'ev}\ and\ \citenamefont
  {Svistunov}(2008{\natexlab{b}})}]{prokofev08}%
  \BibitemOpen
  \bibfield  {author} {\bibinfo {author} {\bibfnamefont {N.~V.}\ \bibnamefont
  {Prokof'ev}}\ and\ \bibinfo {author} {\bibfnamefont {B.~V.}\ \bibnamefont
  {Svistunov}},\ }\bibfield  {title} {\enquote {\bibinfo {title} {{Bold
  Diagrammatic Monte Carlo: A Generic Sign-Problem Tolerant Technique for
  Polaron Models and Possibly Interacting Many-Body Problems}},}\ }\href
  {\doibase 10.1103/PhysRevB.77.125101} {\bibfield  {journal} {\bibinfo
  {journal} {Phys. Rev. B}\ }\textbf {\bibinfo {volume} {77}},\ \bibinfo
  {pages} {125101} (\bibinfo {year} {2008}{\natexlab{b}})}\BibitemShut
  {NoStop}%
\bibitem [{\citenamefont {Punk}\ \emph {et~al.}(2009)\citenamefont {Punk},
  \citenamefont {Dumitrescu},\ and\ \citenamefont {Zwerger}}]{punk09}%
  \BibitemOpen
  \bibfield  {author} {\bibinfo {author} {\bibfnamefont {M.}~\bibnamefont
  {Punk}}, \bibinfo {author} {\bibfnamefont {P.~T.}\ \bibnamefont
  {Dumitrescu}}, \ and\ \bibinfo {author} {\bibfnamefont {W.}~\bibnamefont
  {Zwerger}},\ }\bibfield  {title} {\enquote {\bibinfo {title}
  {{Polaron-to-Molecule Transition in a Strongly Imbalanced Fermi Gas}},}\
  }\href {\doibase 10.1103/PhysRevA.80.053605} {\bibfield  {journal} {\bibinfo
  {journal} {Phys. Rev. A}\ }\textbf {\bibinfo {volume} {80}},\ \bibinfo
  {pages} {053605} (\bibinfo {year} {2009})}\BibitemShut {NoStop}%
\bibitem [{\citenamefont {Vlietinck}\ \emph {et~al.}(2013)\citenamefont
  {Vlietinck}, \citenamefont {Ryckebusch},\ and\ \citenamefont
  {Van~Houcke}}]{vlietinck13}%
  \BibitemOpen
  \bibfield  {author} {\bibinfo {author} {\bibfnamefont {J.}~\bibnamefont
  {Vlietinck}}, \bibinfo {author} {\bibfnamefont {J.}~\bibnamefont
  {Ryckebusch}}, \ and\ \bibinfo {author} {\bibfnamefont {K.}~\bibnamefont
  {Van~Houcke}},\ }\bibfield  {title} {\enquote {\bibinfo {title}
  {{Quasiparticle Properties of an Impurity in a Fermi Gas}},}\ }\href
  {\doibase 10.1103/PhysRevB.87.115133} {\bibfield  {journal} {\bibinfo
  {journal} {Phys. Rev. B}\ }\textbf {\bibinfo {volume} {87}},\ \bibinfo
  {pages} {115133} (\bibinfo {year} {2013})}\BibitemShut {NoStop}%
\bibitem [{\citenamefont {Goulko}\ \emph {et~al.}(2016)\citenamefont {Goulko},
  \citenamefont {Mishchenko}, \citenamefont {Prokof'ev},\ and\ \citenamefont
  {Svistunov}}]{goulko16}%
  \BibitemOpen
  \bibfield  {author} {\bibinfo {author} {\bibfnamefont {O.}~\bibnamefont
  {Goulko}}, \bibinfo {author} {\bibfnamefont {A.~S.}\ \bibnamefont
  {Mishchenko}}, \bibinfo {author} {\bibfnamefont {N.}~\bibnamefont
  {Prokof'ev}}, \ and\ \bibinfo {author} {\bibfnamefont {B.}~\bibnamefont
  {Svistunov}},\ }\bibfield  {title} {\enquote {\bibinfo {title} {{Dark
  Continuum in the Spectral Function of the Resonant Fermi Polaron}},}\ }\href
  {\doibase 10.1103/PhysRevA.94.051605} {\bibfield  {journal} {\bibinfo
  {journal} {Phys. Rev. A}\ }\textbf {\bibinfo {volume} {94}},\ \bibinfo
  {pages} {051605} (\bibinfo {year} {2016})}\BibitemShut {NoStop}%
\bibitem [{\citenamefont {Navon}\ \emph {et~al.}(2010)\citenamefont {Navon},
  \citenamefont {Nascimb{\`e}ne}, \citenamefont {Chevy},\ and\ \citenamefont
  {Salomon}}]{navon10}%
  \BibitemOpen
  \bibfield  {author} {\bibinfo {author} {\bibfnamefont {N.}~\bibnamefont
  {Navon}}, \bibinfo {author} {\bibfnamefont {S.}~\bibnamefont
  {Nascimb{\`e}ne}}, \bibinfo {author} {\bibfnamefont {F.}~\bibnamefont
  {Chevy}}, \ and\ \bibinfo {author} {\bibfnamefont {C.}~\bibnamefont
  {Salomon}},\ }\bibfield  {title} {\enquote {\bibinfo {title} {{The Equation
  of State of a Low-Temperature Fermi Gas with Tunable Interactions}},}\ }\href
  {\doibase 10.1126/science.1187582} {\bibfield  {journal} {\bibinfo  {journal}
  {Science}\ }\textbf {\bibinfo {volume} {328}},\ \bibinfo {pages} {729--732}
  (\bibinfo {year} {2010})}\BibitemShut {NoStop}%
\bibitem [{\citenamefont {Schirotzek}\ \emph {et~al.}(2009)\citenamefont
  {Schirotzek}, \citenamefont {Wu}, \citenamefont {Sommer},\ and\ \citenamefont
  {Zwierlein}}]{schirotzek09}%
  \BibitemOpen
  \bibfield  {author} {\bibinfo {author} {\bibfnamefont {A.}~\bibnamefont
  {Schirotzek}}, \bibinfo {author} {\bibfnamefont {C.-H.}\ \bibnamefont {Wu}},
  \bibinfo {author} {\bibfnamefont {A.}~\bibnamefont {Sommer}}, \ and\ \bibinfo
  {author} {\bibfnamefont {M.~W.}\ \bibnamefont {Zwierlein}},\ }\bibfield
  {title} {\enquote {\bibinfo {title} {{Observation of Fermi Polarons in a
  Tunable Fermi Liquid of Ultracold Atoms}},}\ }\href {\doibase
  10.1103/PhysRevLett.102.230402} {\bibfield  {journal} {\bibinfo  {journal}
  {Phys. Rev. Lett.}\ }\textbf {\bibinfo {volume} {102}},\ \bibinfo {pages}
  {230402} (\bibinfo {year} {2009})}\BibitemShut {NoStop}%
\bibitem [{\citenamefont {Kohstall}\ \emph {et~al.}(2012)\citenamefont
  {Kohstall}, \citenamefont {Zaccanti}, \citenamefont {Jag}, \citenamefont
  {Trenkwalder}, \citenamefont {Massignan}, \citenamefont {Bruun},
  \citenamefont {Schreck},\ and\ \citenamefont {Grimm}}]{kohstall12}%
  \BibitemOpen
  \bibfield  {author} {\bibinfo {author} {\bibfnamefont {C.}~\bibnamefont
  {Kohstall}}, \bibinfo {author} {\bibfnamefont {M.}~\bibnamefont {Zaccanti}},
  \bibinfo {author} {\bibfnamefont {M.}~\bibnamefont {Jag}}, \bibinfo {author}
  {\bibfnamefont {A.}~\bibnamefont {Trenkwalder}}, \bibinfo {author}
  {\bibfnamefont {P.}~\bibnamefont {Massignan}}, \bibinfo {author}
  {\bibfnamefont {G.~M.}\ \bibnamefont {Bruun}}, \bibinfo {author}
  {\bibfnamefont {F.}~\bibnamefont {Schreck}}, \ and\ \bibinfo {author}
  {\bibfnamefont {R.}~\bibnamefont {Grimm}},\ }\bibfield  {title} {\enquote
  {\bibinfo {title} {{Metastability and Coherence of Repulsive Polarons in a
  Strongly Interacting Fermi Mixture}},}\ }\href
  {http://dx.doi.org/10.1038/nature11065} {\bibfield  {journal} {\bibinfo
  {journal} {Nature}\ }\textbf {\bibinfo {volume} {485}},\ \bibinfo {pages}
  {615} (\bibinfo {year} {2012})}\BibitemShut {NoStop}%
\bibitem [{\citenamefont {Scazza}\ \emph {et~al.}(2017)\citenamefont {Scazza},
  \citenamefont {Valtolina}, \citenamefont {Massignan}, \citenamefont {Recati},
  \citenamefont {Amico}, \citenamefont {Burchianti}, \citenamefont {Fort},
  \citenamefont {Inguscio}, \citenamefont {Zaccanti},\ and\ \citenamefont
  {Roati}}]{scazza17}%
  \BibitemOpen
  \bibfield  {author} {\bibinfo {author} {\bibfnamefont {F.}~\bibnamefont
  {Scazza}}, \bibinfo {author} {\bibfnamefont {G.}~\bibnamefont {Valtolina}},
  \bibinfo {author} {\bibfnamefont {P.}~\bibnamefont {Massignan}}, \bibinfo
  {author} {\bibfnamefont {A.}~\bibnamefont {Recati}}, \bibinfo {author}
  {\bibfnamefont {A.}~\bibnamefont {Amico}}, \bibinfo {author} {\bibfnamefont
  {A.}~\bibnamefont {Burchianti}}, \bibinfo {author} {\bibfnamefont
  {C.}~\bibnamefont {Fort}}, \bibinfo {author} {\bibfnamefont {M.}~\bibnamefont
  {Inguscio}}, \bibinfo {author} {\bibfnamefont {M.}~\bibnamefont {Zaccanti}},
  \ and\ \bibinfo {author} {\bibfnamefont {G.}~\bibnamefont {Roati}},\
  }\bibfield  {title} {\enquote {\bibinfo {title} {{Repulsive Fermi Polarons in
  a Resonant Mixture of Ultracold $^{6}\mathrm{Li}$ Atoms}},}\ }\href {\doibase
  10.1103/PhysRevLett.118.083602} {\bibfield  {journal} {\bibinfo  {journal}
  {Phys. Rev. Lett.}\ }\textbf {\bibinfo {volume} {118}},\ \bibinfo {pages}
  {083602} (\bibinfo {year} {2017})}\BibitemShut {NoStop}%
\bibitem [{\citenamefont {Nascimb\`ene}\ \emph {et~al.}(2009)\citenamefont
  {Nascimb\`ene}, \citenamefont {Navon}, \citenamefont {Jiang}, \citenamefont
  {Tarruell}, \citenamefont {Teichmann}, \citenamefont {McKeever},
  \citenamefont {Chevy},\ and\ \citenamefont {Salomon}}]{nascimbene09}%
  \BibitemOpen
  \bibfield  {author} {\bibinfo {author} {\bibfnamefont {S.}~\bibnamefont
  {Nascimb\`ene}}, \bibinfo {author} {\bibfnamefont {N.}~\bibnamefont {Navon}},
  \bibinfo {author} {\bibfnamefont {K.~J.}\ \bibnamefont {Jiang}}, \bibinfo
  {author} {\bibfnamefont {L.}~\bibnamefont {Tarruell}}, \bibinfo {author}
  {\bibfnamefont {M.}~\bibnamefont {Teichmann}}, \bibinfo {author}
  {\bibfnamefont {J.}~\bibnamefont {McKeever}}, \bibinfo {author}
  {\bibfnamefont {F.}~\bibnamefont {Chevy}}, \ and\ \bibinfo {author}
  {\bibfnamefont {C.}~\bibnamefont {Salomon}},\ }\bibfield  {title} {\enquote
  {\bibinfo {title} {{Collective Oscillations of an Imbalanced Fermi Gas: Axial
  Compression Modes and Polaron Effective Mass}},}\ }\href {\doibase
  10.1103/PhysRevLett.103.170402} {\bibfield  {journal} {\bibinfo  {journal}
  {Phys. Rev. Lett.}\ }\textbf {\bibinfo {volume} {103}},\ \bibinfo {pages}
  {170402} (\bibinfo {year} {2009})}\BibitemShut {NoStop}%
\bibitem [{\citenamefont {Combescot}\ \emph {et~al.}(2009)\citenamefont
  {Combescot}, \citenamefont {Giraud},\ and\ \citenamefont
  {Leyronas}}]{combescot09}%
  \BibitemOpen
  \bibfield  {author} {\bibinfo {author} {\bibfnamefont {R.}~\bibnamefont
  {Combescot}}, \bibinfo {author} {\bibfnamefont {S.}~\bibnamefont {Giraud}}, \
  and\ \bibinfo {author} {\bibfnamefont {X.}~\bibnamefont {Leyronas}},\
  }\bibfield  {title} {\enquote {\bibinfo {title} {{Analytical Theory of the
  Dressed Bound State in Highly Polarized Fermi Gases}},}\ }\href
  {http://stacks.iop.org/0295-5075/88/i=6/a=60007} {\bibfield  {journal}
  {\bibinfo  {journal} {EPL}\ }\textbf {\bibinfo {volume} {88}},\ \bibinfo
  {pages} {60007} (\bibinfo {year} {2009})}\BibitemShut {NoStop}%
\bibitem [{\citenamefont {Pilati}\ and\ \citenamefont
  {Giorgini}(2008)}]{pilati08}%
  \BibitemOpen
  \bibfield  {author} {\bibinfo {author} {\bibfnamefont {S.}~\bibnamefont
  {Pilati}}\ and\ \bibinfo {author} {\bibfnamefont {S.}~\bibnamefont
  {Giorgini}},\ }\bibfield  {title} {\enquote {\bibinfo {title} {{Phase
  Separation in a Polarized Fermi Gas at Zero Temperature}},}\ }\href {\doibase
  10.1103/PhysRevLett.100.030401} {\bibfield  {journal} {\bibinfo  {journal}
  {Phys. Rev. Lett.}\ }\textbf {\bibinfo {volume} {100}},\ \bibinfo {pages}
  {030401} (\bibinfo {year} {2008})}\BibitemShut {NoStop}%
\bibitem [{\citenamefont {Recati}\ and\ \citenamefont
  {Stringari}(2010)}]{recati10}%
  \BibitemOpen
  \bibfield  {author} {\bibinfo {author} {\bibfnamefont {A.}~\bibnamefont
  {Recati}}\ and\ \bibinfo {author} {\bibfnamefont {S.}~\bibnamefont
  {Stringari}},\ }\bibfield  {title} {\enquote {\bibinfo {title} {{Spin
  Oscillations of the Normal Polarized Fermi Gas at Unitarity}},}\ }\href
  {\doibase 10.1103/PhysRevA.82.013635} {\bibfield  {journal} {\bibinfo
  {journal} {Phys. Rev. A}\ }\textbf {\bibinfo {volume} {82}},\ \bibinfo
  {pages} {013635} (\bibinfo {year} {2010})}\BibitemShut {NoStop}%
\bibitem [{\citenamefont {Riedl}\ \emph {et~al.}(2008)\citenamefont {Riedl},
  \citenamefont {S\'anchez~Guajardo}, \citenamefont {Kohstall}, \citenamefont
  {Altmeyer}, \citenamefont {Wright}, \citenamefont {Denschlag}, \citenamefont
  {Grimm}, \citenamefont {Bruun},\ and\ \citenamefont {Smith}}]{riedl08}%
  \BibitemOpen
  \bibfield  {author} {\bibinfo {author} {\bibfnamefont {S.}~\bibnamefont
  {Riedl}}, \bibinfo {author} {\bibfnamefont {E.~R.}\ \bibnamefont
  {S\'anchez~Guajardo}}, \bibinfo {author} {\bibfnamefont {C.}~\bibnamefont
  {Kohstall}}, \bibinfo {author} {\bibfnamefont {A.}~\bibnamefont {Altmeyer}},
  \bibinfo {author} {\bibfnamefont {M.~J.}\ \bibnamefont {Wright}}, \bibinfo
  {author} {\bibfnamefont {J.~Hecker}\ \bibnamefont {Denschlag}}, \bibinfo
  {author} {\bibfnamefont {R.}~\bibnamefont {Grimm}}, \bibinfo {author}
  {\bibfnamefont {G.~M.}\ \bibnamefont {Bruun}}, \ and\ \bibinfo {author}
  {\bibfnamefont {H.}~\bibnamefont {Smith}},\ }\bibfield  {title} {\enquote
  {\bibinfo {title} {{Collective Oscillations of a Fermi Gas in the Unitarity
  Limit: Temperature Effects and the Role of Pair Correlations}},}\ }\href
  {\doibase 10.1103/PhysRevA.78.053609} {\bibfield  {journal} {\bibinfo
  {journal} {Phys. Rev. A}\ }\textbf {\bibinfo {volume} {78}},\ \bibinfo
  {pages} {053609} (\bibinfo {year} {2008})}\BibitemShut {NoStop}%
\bibitem [{\citenamefont {Chiacchiera}\ \emph {et~al.}(2009)\citenamefont
  {Chiacchiera}, \citenamefont {Lepers}, \citenamefont {Davesne},\ and\
  \citenamefont {Urban}}]{chiacchiera09}%
  \BibitemOpen
  \bibfield  {author} {\bibinfo {author} {\bibfnamefont {S.}~\bibnamefont
  {Chiacchiera}}, \bibinfo {author} {\bibfnamefont {T.}~\bibnamefont {Lepers}},
  \bibinfo {author} {\bibfnamefont {D.}~\bibnamefont {Davesne}}, \ and\
  \bibinfo {author} {\bibfnamefont {M.}~\bibnamefont {Urban}},\ }\bibfield
  {title} {\enquote {\bibinfo {title} {{Collective Modes of Trapped Fermi Gases
  With In-Medium Interaction}},}\ }\href {\doibase 10.1103/PhysRevA.79.033613}
  {\bibfield  {journal} {\bibinfo  {journal} {Phys. Rev. A}\ }\textbf {\bibinfo
  {volume} {79}},\ \bibinfo {pages} {033613} (\bibinfo {year}
  {2009})}\BibitemShut {NoStop}%
\bibitem [{\citenamefont {Lepers}\ \emph {et~al.}(2010)\citenamefont {Lepers},
  \citenamefont {Davesne}, \citenamefont {Chiacchiera},\ and\ \citenamefont
  {Urban}}]{lepers10}%
  \BibitemOpen
  \bibfield  {author} {\bibinfo {author} {\bibfnamefont {T.}~\bibnamefont
  {Lepers}}, \bibinfo {author} {\bibfnamefont {D.}~\bibnamefont {Davesne}},
  \bibinfo {author} {\bibfnamefont {S.}~\bibnamefont {Chiacchiera}}, \ and\
  \bibinfo {author} {\bibfnamefont {M.}~\bibnamefont {Urban}},\ }\bibfield
  {title} {\enquote {\bibinfo {title} {{Numerical Solution of the Boltzmann
  Equation for the Collective Modes of Trapped Fermi Gases}},}\ }\href
  {\doibase 10.1103/PhysRevA.82.023609} {\bibfield  {journal} {\bibinfo
  {journal} {Phys. Rev. A}\ }\textbf {\bibinfo {volume} {82}},\ \bibinfo
  {pages} {023609} (\bibinfo {year} {2010})}\BibitemShut {NoStop}%
\bibitem [{\citenamefont {Chiacchiera}\ \emph {et~al.}(2011)\citenamefont
  {Chiacchiera}, \citenamefont {Lepers}, \citenamefont {Davesne},\ and\
  \citenamefont {Urban}}]{chiacchiera11}%
  \BibitemOpen
  \bibfield  {author} {\bibinfo {author} {\bibfnamefont {S.}~\bibnamefont
  {Chiacchiera}}, \bibinfo {author} {\bibfnamefont {T.}~\bibnamefont {Lepers}},
  \bibinfo {author} {\bibfnamefont {D.}~\bibnamefont {Davesne}}, \ and\
  \bibinfo {author} {\bibfnamefont {M.}~\bibnamefont {Urban}},\ }\bibfield
  {title} {\enquote {\bibinfo {title} {{Role of Fourth-Order Phase-Space
  Moments in Collective Modes of Trapped Fermi Gases}},}\ }\href {\doibase
  10.1103/PhysRevA.84.043634} {\bibfield  {journal} {\bibinfo  {journal} {Phys.
  Rev. A}\ }\textbf {\bibinfo {volume} {84}},\ \bibinfo {pages} {043634}
  (\bibinfo {year} {2011})}\BibitemShut {NoStop}%
\bibitem [{\citenamefont {Pantel}\ \emph {et~al.}(2012)\citenamefont {Pantel},
  \citenamefont {Davesne}, \citenamefont {Chiacchiera},\ and\ \citenamefont
  {Urban}}]{pantel12}%
  \BibitemOpen
  \bibfield  {author} {\bibinfo {author} {\bibfnamefont {P.-A.}\ \bibnamefont
  {Pantel}}, \bibinfo {author} {\bibfnamefont {D.}~\bibnamefont {Davesne}},
  \bibinfo {author} {\bibfnamefont {S.}~\bibnamefont {Chiacchiera}}, \ and\
  \bibinfo {author} {\bibfnamefont {M.}~\bibnamefont {Urban}},\ }\bibfield
  {title} {\enquote {\bibinfo {title} {{Trap Anharmonicity and Sloshing Mode of
  a Fermi Gas}},}\ }\href {\doibase 10.1103/PhysRevA.86.023635} {\bibfield
  {journal} {\bibinfo  {journal} {Phys. Rev. A}\ }\textbf {\bibinfo {volume}
  {86}},\ \bibinfo {pages} {023635} (\bibinfo {year} {2012})}\BibitemShut
  {NoStop}%
\bibitem [{\citenamefont {Chiacchiera}\ \emph {et~al.}(2013)\citenamefont
  {Chiacchiera}, \citenamefont {Davesne}, \citenamefont {Enss},\ and\
  \citenamefont {Urban}}]{chiacchiera13}%
  \BibitemOpen
  \bibfield  {author} {\bibinfo {author} {\bibfnamefont {S.}~\bibnamefont
  {Chiacchiera}}, \bibinfo {author} {\bibfnamefont {S.}~\bibnamefont
  {Davesne}}, \bibinfo {author} {\bibfnamefont {T.}~\bibnamefont {Enss}}, \
  and\ \bibinfo {author} {\bibfnamefont {M.}~\bibnamefont {Urban}},\ }\bibfield
   {title} {\enquote {\bibinfo {title} {{Damping of the Quadrupole Mode in a
  Two-Dimensional Fermi Gas}},}\ }\href {\doibase 10.1103/PhysRevA.88.053616}
  {\bibfield  {journal} {\bibinfo  {journal} {Phys. Rev. A}\ }\textbf {\bibinfo
  {volume} {88}},\ \bibinfo {pages} {053616} (\bibinfo {year}
  {2013})}\BibitemShut {NoStop}%
\bibitem [{\citenamefont {Landau}\ and\ \citenamefont
  {Lifshitz}(1966)}]{landau66}%
  \BibitemOpen
  \bibfield  {author} {\bibinfo {author} {\bibfnamefont {L.~D.}\ \bibnamefont
  {Landau}}\ and\ \bibinfo {author} {\bibfnamefont {E.~M.}\ \bibnamefont
  {Lifshitz}},\ }\href@noop {} {\emph {\bibinfo {title} {Course of Theoretical
  Physics, Vol. 6: Fluid Mechanics}}}\ (\bibinfo  {publisher}
  {Butterworth-Heinemann},\ \bibinfo {year} {1966})\BibitemShut {NoStop}%
\bibitem [{\citenamefont {Vichi}(2000)}]{vichi00}%
  \BibitemOpen
  \bibfield  {author} {\bibinfo {author} {\bibfnamefont {L.}~\bibnamefont
  {Vichi}},\ }\bibfield  {title} {\enquote {\bibinfo {title} {{Collisional
  Damping of the Collective Oscillations of a Trapped Fermi Gas}},}\ }\href
  {\doibase 10.1023/A:1004815907236} {\bibfield  {journal} {\bibinfo  {journal}
  {J. Low T. Phys.}\ }\textbf {\bibinfo {volume} {121}},\ \bibinfo {pages}
  {177} (\bibinfo {year} {2000})}\BibitemShut {NoStop}%
\end{thebibliography}%

\end{document}